\documentclass[%
 aip,
rsi,%
 amsmath,amssymb,
 reprint,%
]{revtex4-1}

\let\oldciteauthor=\citeauthor
\def\citeauthor#1{\hypersetup{citecolor=black}\oldciteauthor{#1}}
\let\oldcite=\cite
\def\cite#1{\hypersetup{citecolor=blue}\oldcite{#1}}

\usepackage{graphicx}

\usepackage{physics}
\usepackage{braket}
\usepackage{nccmath}
\usepackage{tensor}
\usepackage[document]{ragged2e}
\setcitestyle{super}
\usepackage[colorlinks=true,citecolor=blue,linkcolor=blue]{hyperref}
\usepackage{dcolumn}
\usepackage{bm}

\begin{document}

\preprint{AIP/123-QED}

\title{Experimental Simulation of Hybrid Quantum Systems and Entanglement on a Quantum Computer}

\author{Farai Mazhandu}
\affiliation{Nano Scale Transport Physics Laboratory, School of Physics, University of the Witwatersrand, Private Bag 3, WITS 2050, Johannesburg, South Africa \\
}

\author{Kayleigh Mathieson}
\affiliation{Nano Scale Transport Physics Laboratory, School of Physics, University of the Witwatersrand, Private Bag 3, WITS 2050, Johannesburg, South Africa \\
}

\author{Christopher Coleman}
\affiliation{Nano Scale Transport Physics Laboratory, School of Physics, University of the Witwatersrand, Private Bag 3, WITS 2050, Johannesburg, South Africa \\
}

\author{Somnath Bhattacharyya}
\email{somnath.bhattacharyya@wits.co.za}
\affiliation{Nano Scale Transport Physics Laboratory, School of Physics, University of the Witwatersrand, Private Bag 3, WITS 2050, Johannesburg, South Africa \\
}
\affiliation{National University of Science and Technology “MISiS”, Leninski Avenue 4, 119991 Moscow, Russia \\
}

\date{3 September 2019}

\begin{abstract}
\small{We propose the utilization of the IBM Quantum Experience quantum computing system to simulate different scenarios involving common hybrid quantum system components, the Nitrogen Vacancy Centre (NV centre) and the Flux Qubit. We perform a series of the simulation experiments and demonstrate properties of a virtual hybrid system, including its spin relaxation rate and state coherence. In correspondence with experimental investigations we look at the scalability of such systems and show that increasing the number of coupled NV centres decreases the coherence time. We also establish the main error rate as a function of the number of control pulses in evaluating the fidelity of the four qubit virtual circuit with the simulator. Our results show that the virtual system can attain decoherence and fidelity values comparable to what has been reported for experimental investigations of similar physical hybrid systems, observing a coherence time at $0.35$ s for a single NV centre qubit and fidelity in the range of $0.82$. The work thus establishes an effective simulation test protocol for different technologies to test and analyze them before experimental investigations or as a supplementary measure.}
\end{abstract}

\maketitle

\small

Quantum computers have the potential to solve problems that scale up at polynomial time and are thus predicted to outperform classical computers in a wide range of tasks including machine learning\cite{Biamonte2017}, complex simulations\cite{RevModPhys.86.153,Lesovik2019,Zhukov2017ModelingDO,GarciaMartin2018FiveET,PhysRevA.99.022308,Raeisi2012,Smith:2019mek,Hegade2017ExperimentalDO,Kapil2018QuantumSO,Gerritsma2010QuantumSO,PhysRevA.99.062122,Chang,Yan753} and optimization problems\cite{Moll_2018}. However, to establish true quantum supremacy, there is a need to build and demonstrate universal fault tolerant and scalable computing systems that can extend beyond the capabilities of classical computational systems\cite{Chao2018}. To accomplish this several different types of systems and architectures have been proposed and continue to be studied.

\smallbreak

More recently this has included the demonstration of hybrid systems which combine different complementary quantum device elements, often coupling a combination of a superconducting, atomic and or spin systems into a single  circuit\cite{Kurizki3866,PhysRevLett.107.220501}. NV centres have been studied extensively for this purpose, this is because the spin states associated with the NV centre present a well-studied energy level splitting which can be readily accessed and addressed through both electrical and optical measurement techniques and are thus able to couple relatively easily to circuitry and other device components. It has already been shown that coupled NV centres and flux qubits\cite{PhysRevLett.105.210501,Qiu2014AHQ} are ideal complimentary elements for such hybrid systems. As both these device elements rely on spin they can easily be coupled, and experimental investigations have demonstrated quantum information transfer between flux qubit and NV centre ensembles\cite{PhysRevLett.105.210501}. Additionally, NV centres present ideal quantum logic elements and have shown to be useful for a range of operations including as quantum registers and as quantum gates\cite{VanderSar2012,Taminiau2014}. Due to their possibility of realizing fault tolerant logic operations holonomic quantum gates have been widely studied in various systems \cite{VanderSar2012,Taminiau2014,Zu2014,Nagata2018,PhysRevLett.110.190501,Li2017,Zhou2017} but most notably with the geometric phase in NV centre qubits \cite{Zhukov2017ModelingDO,GarciaMartin2018FiveET,PhysRevA.99.022308}. 

\smallbreak

We attempt to extend the work of \citeauthor{PhysRevA.94.032329} \cite{PhysRevA.94.032329} which suggested the possibility of simulating different quantum systems, their states and properties with the IBM Quantum Experience cloud-based computing platform. Our work investigates the physical properties of three variants of a small-scale circuit onto which we map an equivalent virtual system composed of the NV centre(s) and a flux qubit. We relate this to the work of \citeauthor{Zhukov2017ModelingDO} \cite{Zhukov2017ModelingDO}, that looks at simulating spin qubits with the IBM Quantum Experience (5 qubit) IBMqx4. Also, we draw inspiration from the work of \citeauthor{PhysRevB.89.045432} \cite{PhysRevB.89.045432}, which makes use of the properties of NV centres and their topological transitions to act as a simulation platform itself; which we instead extend for the purpose of simulating their useful properties for quantum circuits. 

\smallbreak

Firstly, we examine the relaxation rate and state coherence of a three logical qubit entangled circuit. By tuning the initial state preparation through microwave pulse control of the logic operations of the simulator, we are able to simulate dynamics equivalent to that observed in the entangled electron, nitrogen atom and instead replace the ${}^{13}C$ atom with a flux qubit, which relates to what is found in physical NV centre systems. Next, we use the same technique to simulate a coupled system involving three NV centres and a flux qubit and demonstrate effects in accordance to recent theoretical predictions\cite{PhysRevB.74.161203}. To demonstrate the versatility of this simulation scheme we also investigated the effects of scaling up the system and performing a complete circuit simulation by increasing the number of NV centres coupled to the flux qubit, and modelling their dissipation according to an Ornstein-Uhlenbeck process\cite{Jing2018}. 

\smallbreak

\begin{figure}
    \centering
    \includegraphics[height=3cm, width=8cm]{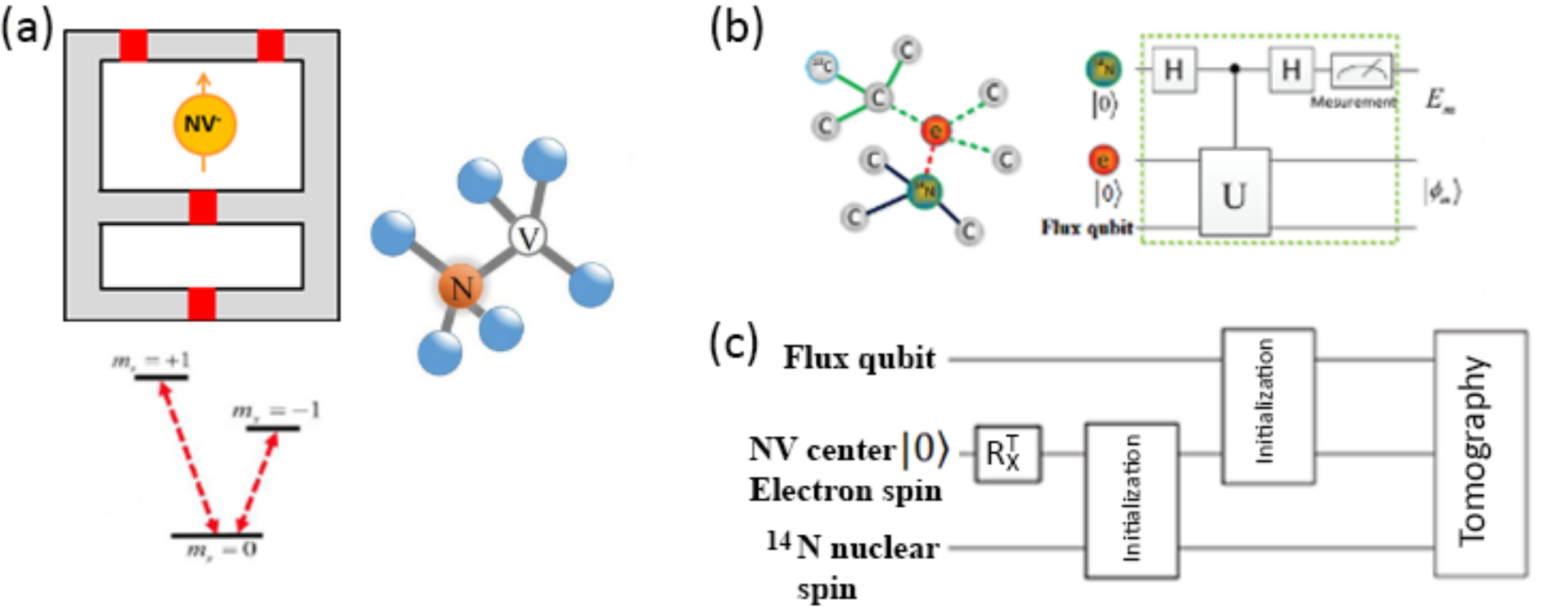}
    \caption{(a) NV centre coupled to a superconducting flux qubit with four Josephson junctions (in red), crystal-lattice and energy diagrams of an NV center in diamond. (b) Nitrogen vacancy center in diamond and a representation of the unitary decomposition on the circuit line. (c) Two qubit control and flux qubit-nuclear entangling gate structure.}
    \label{fig:figure1}
\end{figure}

As shown in Figure 1, the first simulation experiment involves a direct investigation of the physical components that collectively make up a single nitrogen vacancy centre: the electron spin coupled to the nuclear spin of its nitrogen atom and the nuclear spin of a nearby $^{13}C$ atom. This diamond lattice defect has five valence electrons, as well as an extra electron captured from the environment forming a negatively charged state. These six electrons occupy the molecular levels forming a spin-triplet (S = 1) state. In experiments, applying an external magnetic field can control the energy level splitting through a Zeeman effect $m_s$ = 0 or $\pm$ 1 spin, which can form a two-level system $m_s$ = 1 and $m_s$ = 0. It is well established that appreciable quantities of the natural abundance ${}^{13}C$ isotopes can act as a source of decoherence of the NV centre through spin-spin correlations. It is thus interesting to incorporate such elements into the simulation experiments, as such dynamics can be easily compared to experiments\cite{Childress281,PhysRevLett.102.057403}. The Hamiltonian of the two spins and flux qubit in an external magnetic field $B_{0}$ is ($\hbar$ = 1):

\begin{align}
    H = &DS_z^2 + \gamma_e B_0 S_z - \gamma_{N} B_0 N_z - B_{0} \bigg(\frac{\delta}{2} \sigma_{x}\bigg) \nonumber\\ 
    &+ Q \bigg(\frac{\delta}{2} \sigma_{x}\bigg)^{2} - \frac{\delta}{2} \sigma_{x} + J_C S_z g_f \sigma_{z} + J_N S_z  N_z
    \label{fig:eq1}
\end{align}

Here $S_z$ and $N_z$ are operators for the electron spin and the $^{14}$N nuclear spin. In the MHz and $2\pi$ basis with $D$ being the electron spin zero-field splitting tensor from the anistropic magnetic dipole-dipole interaction is taken as being $2.87$ $\times$ $10^{3}$. We take the electron $g$-factors $\gamma_e$ = $2.8$ and $\gamma_N$ = $0.3077$ $\times$ $10^{-3}$. The hyperfine coupling constant between the electron spin and flux qubit is taken as $J_C$ = 14. $Q$ = $-5.1$ is the quadrupole splitting tensor of the nitrogen nuclear spin. The hyperfine coupling constant between the electron spin and the $^{14}$N nuclear spin is $J_N$ = 2.1. We take the term for the flux qubit $- \frac{\delta}{2} \sigma_{x}$, and the term for the flux qubit interaction $g_f \sigma_{z}$. Where we take $-\frac{\delta}{2}$ to be the flux tunable parameter, the Pauli matrices $\sigma_{x}$ and $\sigma_{z}$, and $g_f$ a suitable coupling parameter between the flux qubit and the NV centres.

\begin{figure}
    \centering
    \includegraphics[height=2.8cm, width=8cm]{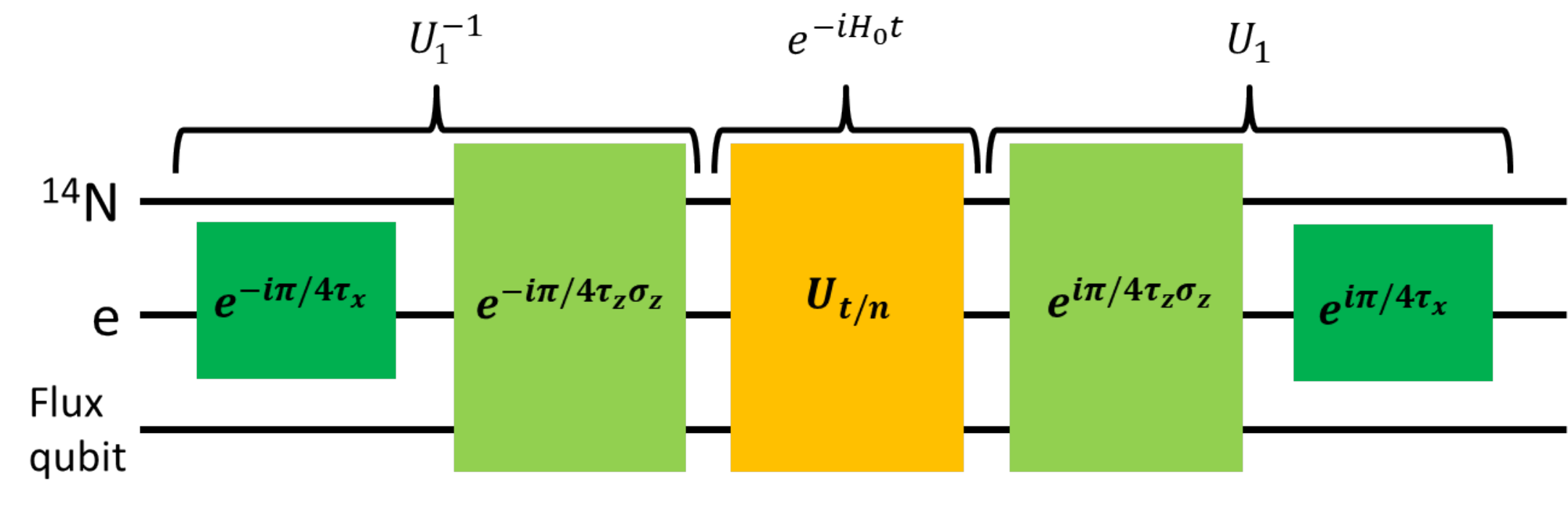}
    \caption{Quantum circuit for the unitary time evolution of $e^{-i H_0 t}$ on the $^{14}$N, the electron spin $e$ and the flux qubit acting as qubits. We apply the Trotter approximation to $U_1$.}
    \label{fig:figure2}
\end{figure}

\begin{figure}
    \centering
    \includegraphics[height=2cm, width=8cm]{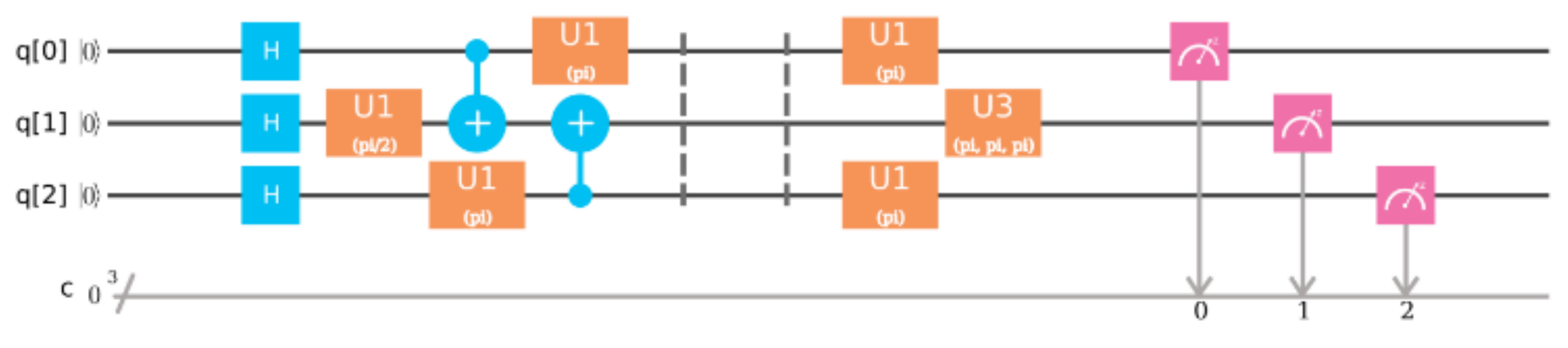}
    \caption{Main three qubit quantum circuit on IBM Q consisting of Hadamard gates (H), controlled-not (CNOT) gates, unitary gates $U_1$ and $U_3$ (being the inverse), as well respective measurements.}
    \label{fig:IBMcircuit1}
\end{figure}

\smallbreak

The IBM Quantum Experience’s IBMqx4 has already demonstrated its potential to be utilized by for quantum simulations\cite{PhysRevA.94.032329} of this scale. In our simulations the virtual system composed of individual two-level spin states of electron, nitrogen and the flux qubit are represented by three separate qubit lines on the physical circuit. We designate the electron-spin line for control and readout of the entangled components. This allows for easy comparison to recent experiments where electron-spin in nitrogen doped diamond were used as multi-qubit spin registers. We are mainly concerned with logic protocols that involve the entanglement of the electron-spin line, and thus measurements can only be made in the Z basis on the electron-spin qubit. We take all the qubits in the zero state and map the three qubits on the IBMqx4. In order to do this the Hamiltonian (1) is transformed and mapped into an experimentally realizable Hamiltonian through the Trotter approximation method\cite{Lamata2017}, which we show in the supplementary information.

\smallbreak

In Figure 3 we show the quantum circuit lines with the unitary time evolution and apply the Trotter approximation to $U_{1}$, then in Figure 2 we reflect this on the simulator. Adjusting the external magnetic field adjusts the coupling to different operational modes, in the physical system. NV centre experiments have shown that the decoherence for a single NV centre is $\approx$ 14 ms and electron spin (3 $\mu$s), furthermore isotopically pure diamond has a time of $\approx$ 3 ms\cite{Bar-Gill2013,PhysRevX.8.031025,Herbschleb2019}. 

\smallbreak

We take the $^{14}N$ and $^{13}C$ as a source of noise approximating a Markovian-Gaussian probability distribution governed by a static magnetic field and noise from the electron spins to follow an Ornstein-Uhlenbeck\cite{Jing2018} process controlled by coupling to the bath. A challenge that arises in constructing and simulating open quantum systems is decoherence, as the inter-qubit dynamics that implement the quantum logic are unavoidably affected by uncontrolled couplings to the solid-state environment, preventing high-fidelity gate performance. 

\smallbreak

An error correction technique that has been investigated in systems involving NV centres is dynamical decoupling\cite{PhysRevLett.82.2417,PhysRevLett.121.220502,PhysRevB.85.155204}, which decouples the system from its environmental degrees of freedom. We look at this method in more depth and in specific cases in the supplementary information. Dynamical decoupling has been shown to lengthen their electron spin coherence time \cite{deLange60,Hanson352,PhysRevB.74.161203}; and it allows each qubit to be individually and uniformly decoupled, with the added sequence the flux qubit coherence time ranges from 40 $\mu$s to 85 $\mu$s \cite{Yan2016}. 

\begin{figure}
    \centering
    \includegraphics[height=14cm, width=7cm]{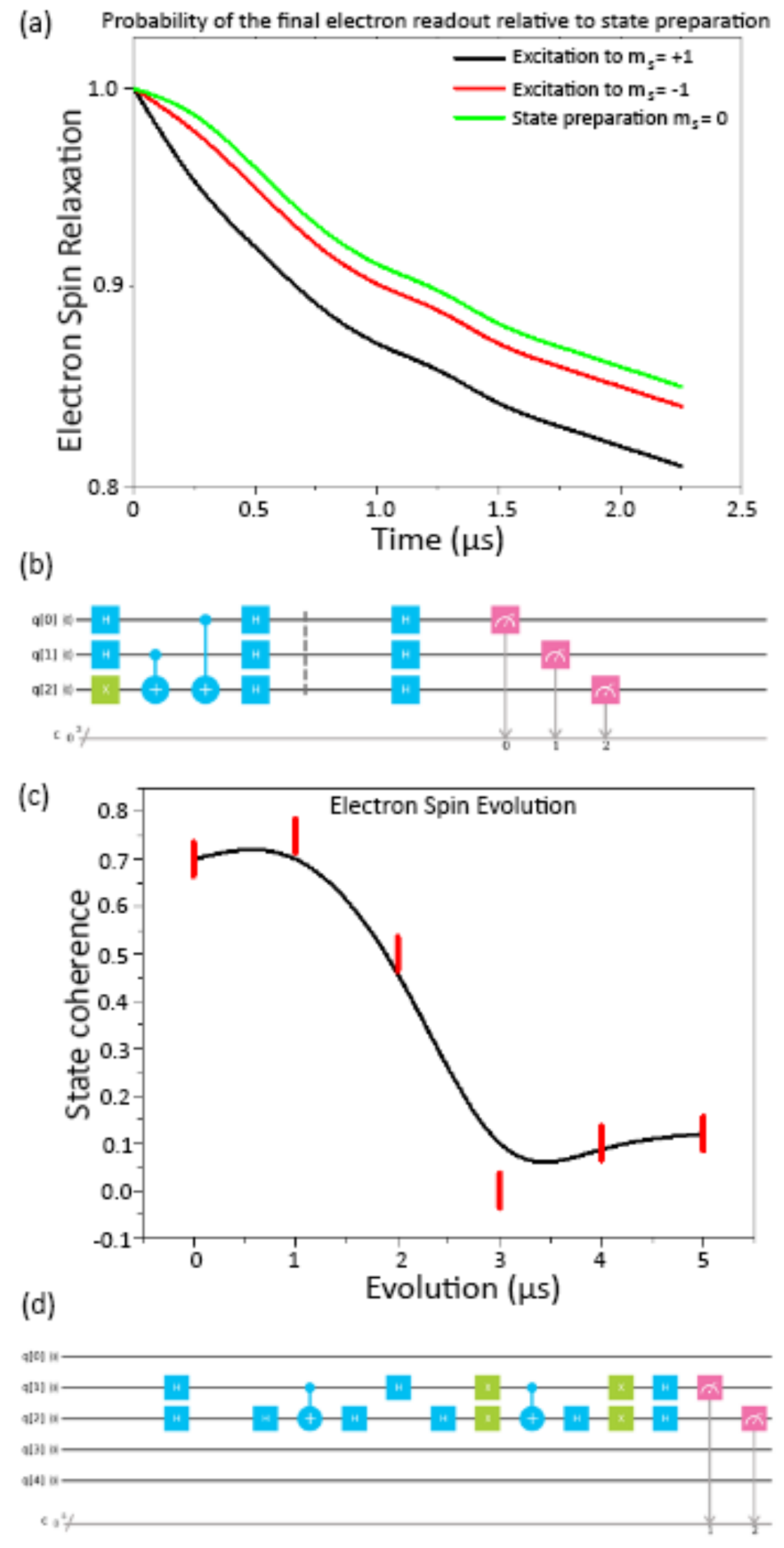}
    \caption{Electron spin relaxation as a function of pulse duration. (b) Decoherence variation with time for entangled electron spin of the NV centre. (c) Quantum circuit used to simulate central spin relaxation. (d) The entangled spin evolution.}
    \label{fig:figure4}
\end{figure}

\begin{figure}
    \centering
    \includegraphics[height=2cm, width=8cm]{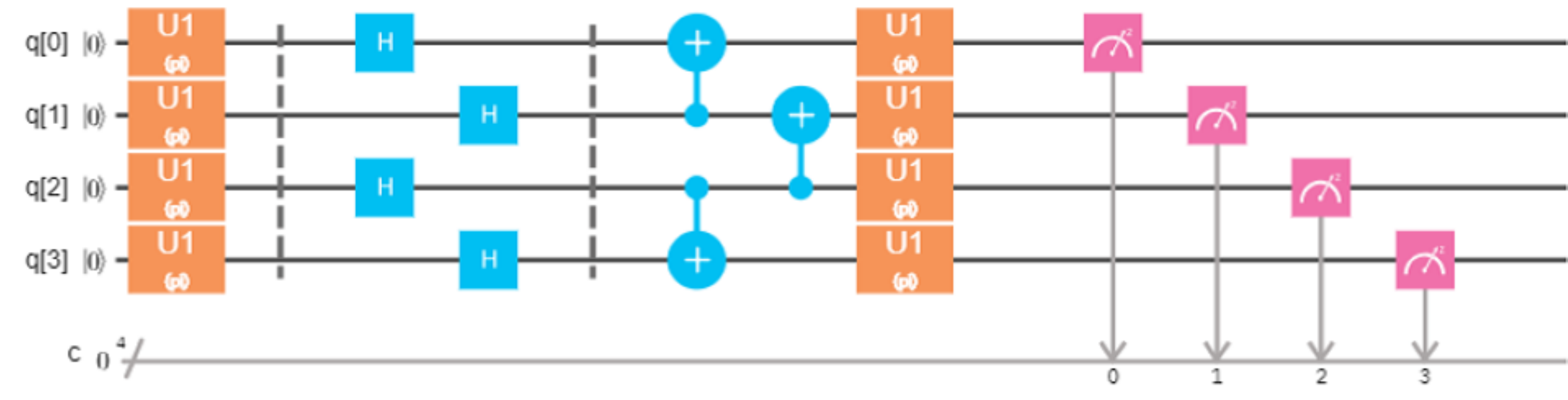}
    \caption{The gate structure for the extended model with $n$ NV centres and a flux qubit.}
    \label{fig:IBMcircuit2}
\end{figure}

\smallbreak

The electron spin of the NV centre is always coupled to the nuclear spin of its own nitrogen atom, we show this in Figure 4 where in (a) we look at the electron spin relaxation in terms of the pulses applied in the simulation process and use (b) as our circuit to simulate this. The state coherence decreases and then stabilizes with increased pulse duration, within the 2.5 $\mu$s time instance it decreases by 0.2. We look at this in terms of the state excitations to $m_s$ = $+1$ and $m_s$ = $-1$, as well as the state preparation to $m_s$ = 0, where we see a consistent relation with the state preparation having a slower decrease in relaxation and the excitation to $m_s$ = $+1$ decreasing fastest. In (c) we look at the electron spin state coherence evolution which we simulate with the gate structure in (d), this exhibits a steady and then proceedingly rapid decline, from time 0 to 1 $\mu$s, then 1 to 3 $\mu$s respectively; it subsequently increases to then stabilize at the 4 to 5 $\mu$s interval. The relaxation observed from our virtual system's electron spin resembles relaxation shown in physical NV center experiments \cite{Bar-Gill2013,Mrozek2015,Schwartzeaat8978}. In physical experiments this is often a result of perturbation by the $^{13}C$ nuclear spins around it which decoheres the spin center, but as the pulse sequence comes into effect the system transitions into a steady state. In physical experiments the longitudinal relaxation observed in experiments has displayed both temperature and magnetic field dependence\cite{Bar-Gill2013,Mrozek2015,Schwartzeaat8978}. The simulator has its limitations, but the relaxation we observe from our virtual system scales similarly to the experimentally verified relaxation rate\cite{Bar-Gill2013}. We relate the decoherence of the virtual system to the reduced state coherence observed in our quantum gate simulation of the system with the simulator. The decoherence affecting the physical IBM Q qubit likely has a small but not overly significant contribution to the time scale and our measurements. This virtual system simulation has similarities to the work done by \citeauthor{Taminiau2014}\cite{Taminiau2014}, which we look at in certain cases in the supplementary information. 

\smallbreak

We now look at a four qubit system with three NV centres and a flux qubit, which is an experimentally realizable hybrid quantum circuit architecture \cite{Song2015}. A technique of entangling the qubits in physical experiments is through using a resonant driving field and letting the system evolve through a rotating unitary operation. This virtual circuit can be represented by a Hamiltonian which is given in the supplementary information. 

\begin{figure}
    \centering
    \includegraphics[height=6cm, width=8cm]{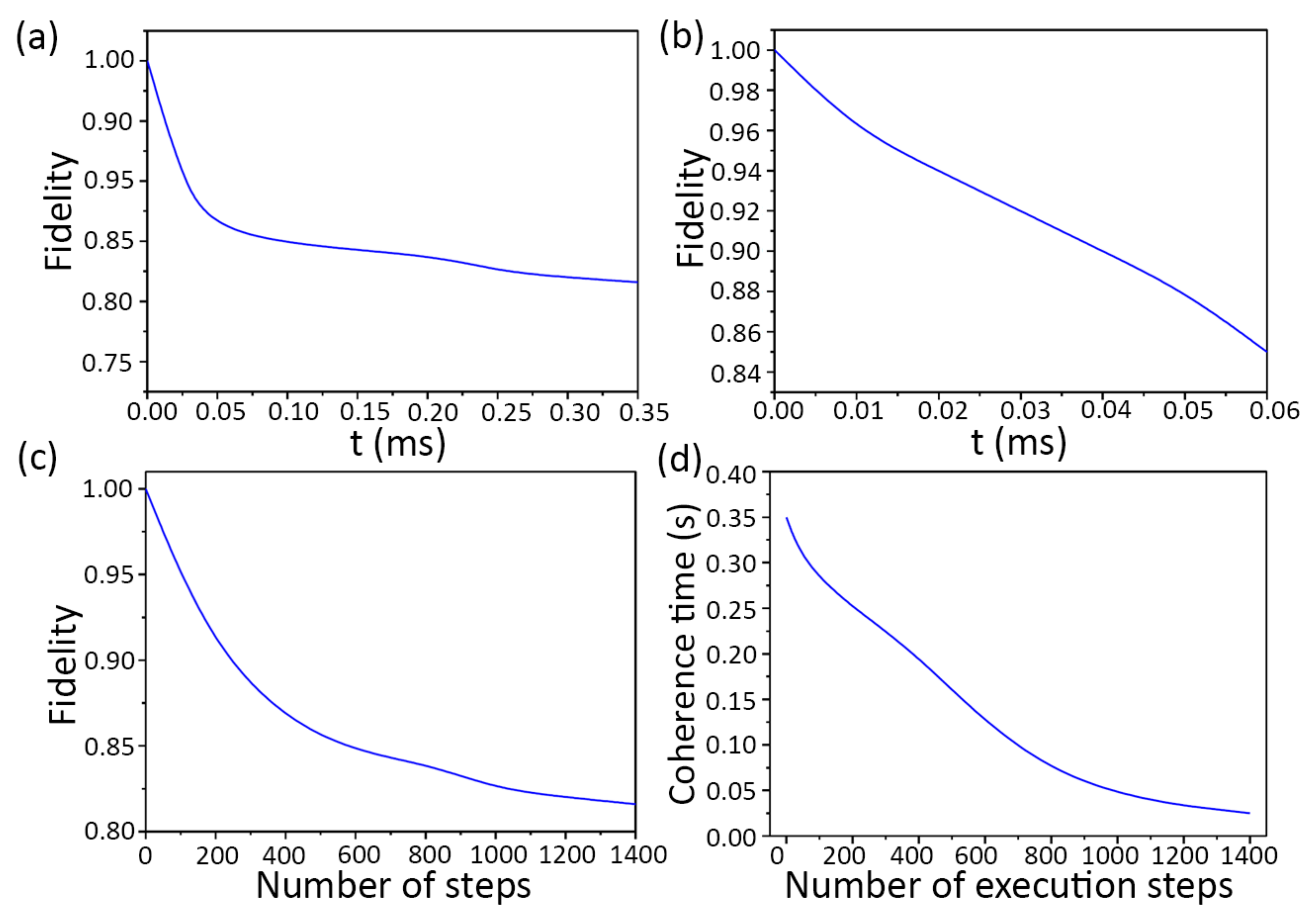}
    \caption{(a) Fidelity of a four qubit system involving three NV centres and a flux qubit. (b) A magnified view of the four qubit fidelity in a shorter time frame (from another run of the simulation). (c) Fidelity of a four qubit system involving three NV centres and a flux qubit vs. number of steps on IBM Q. (d) State coherence vs. number of execution steps}
    \label{fig:figure5}
\end{figure}

\smallbreak

Fidelity which is the measurement of the closeness between two quantum states is measured for our gate simulation. By studying the influence of the shape and power of the control pulses, we can investigate the degree of influence of the IBM quantum system and its noise on our circuit simulation. In Figure 6 (a) we look at a four qubit gate based logical circuit which consists of three NV centres and a flux qubit, here we observe that the fidelity decreases quickly from 0 to 0.5 ms and then slowly to level at 0.82 at 0.35 ms. We consider the initial decline in a more magnified view in (b), which shows the reduction of the state fidelity in 0.06 ms. To show that our results obtained from our virtual system simulations were reasonable, we look at the limits of the simulation platform in performing this circuit simulation by showing the variation in the fidelity and state coherence with varied numbers of steps applied in processing the quantum gates; this correlates to the number of times it is run on IBM's 5 qubit system. In (c) we look at the fidelity for the four qubit system, and find a steady decrease in the fidelity with the number of steps involved. The fidelity eventually stabilizes. This indicates that there is only a limited contribution of the physical IBM Q quantum system and its decoherence and dissipation to the simulation of our virtual system. Similarly in (d) we look at the coherence times which shows a similar, although more prominent decrease in the times observed with an increase in the number of execution steps of the experiment. This decreases to almost 0 at 1400 steps. The results of our circuit simulation are related to results obtained in both theoretical and experimental investigations of similar circuits$\cite{Maleki:18,Lei2017,Liu2016a}$.

\smallbreak

With the system involving $n$ NV centres depicted by Figure 5 we look at how the increase in the number $n$ of affects the coherence time. This is shown in Figure 7, in (a) we show the coherence time scaling linearly with the increase in the number of NV centre qubits, where it decreases by 0.15 s with a change from one to four in number $n$. In (b) we show the change in the decoherence with the evolution time in ms; it illustrates that there isn't a large difference between having one and two NV centres in the system since they scale similarly with only slight differences in their rate of evolution. With three and four qubits they each decohere faster within a shorter evolution time. With numbers $n$ of one to three of NV centres they all decohere to 0.4 at around 27 ms and the system with four NV centres decoheres to 0.4 in 8 ms, which is significantly faster. The state coherence stabilizes as the time approaches 30 ms. This shows the transition to a steady state in the system, which is common to both the simulator and the virtual system at approximately 0.4. 

\smallbreak

IBM Q's superconducting qubits have a coherence time which ranges from 50 - 100 $\mu$s (0.05 - 0.1 ms), as shown in Figures 4, 6 and 7, our simulated system displays a longer coherence time than that reaching around 0.35 s. In both the virtual and experimental circuits each qubit is entangled and the dynamics between the qubits, such as the transfer of excitations can be blocked through adequate phase and coupling parameters. This can lead to constructive and destructive interference effects in the circuit operations. In physical systems the coupling strength is an important factor for quantum memory operation as the time taken to transfer information between the systems is inversely proportional to the coupling strength, so faster operation within its coherence time leads to high-fidelity operation.

\begin{figure}
    \centering
    \includegraphics[height=3cm, width=8cm]{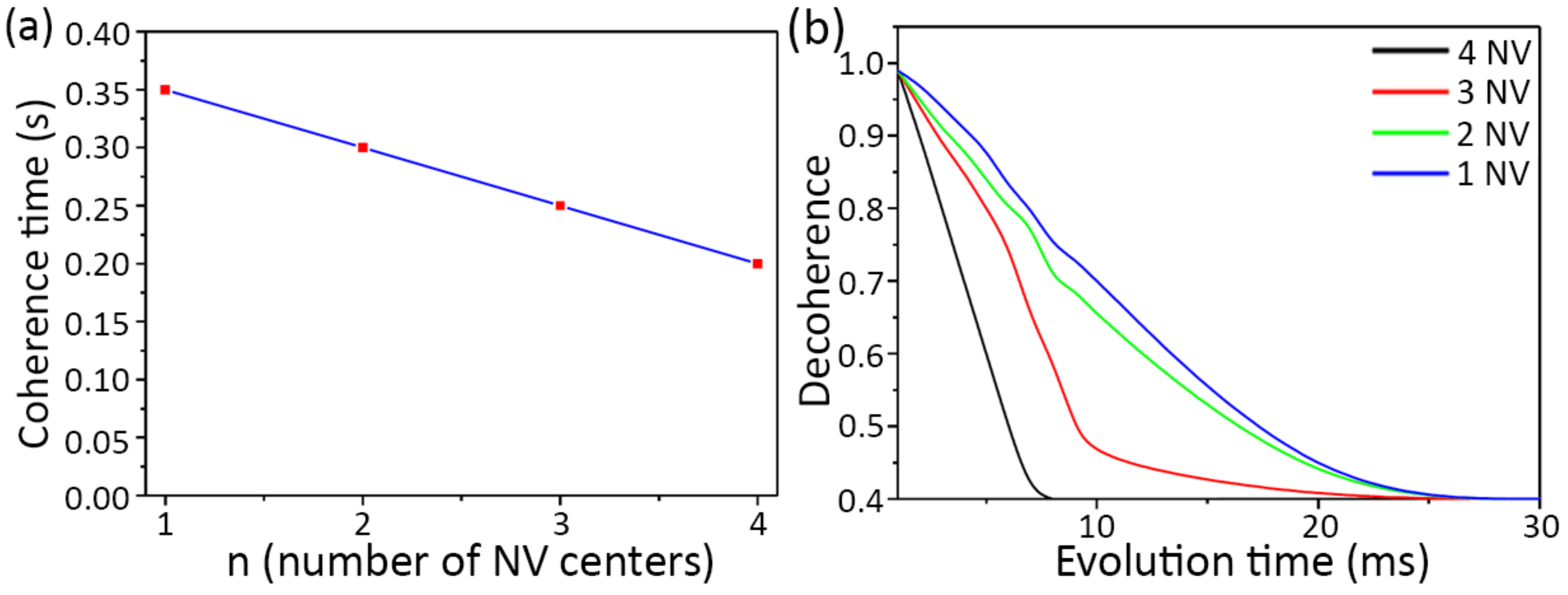}
    \caption{(a) Coherence vs. n (number of NV centres). (b) Decoherence vs. evolution time, with 1 NV (blue), 2 NVs (green), 3 NVs (red) and 4 NVs (black).}
    \label{fig:figure6}
\end{figure}

\smallbreak

In conclusion, we have performed a digital simulation involving control pulses acting on a three different hybrid quantum systems based on NV centres in diamond and a four-junction flux qubit. The decoherence and fidelity patterns observed from our virtual system with the simulator are consistent with theoretical predictions and experimental results for similar systems, which we largely attribute to our usage of a controllable coupling strength and decoupling pulses to reduce noise. This work demonstrates the possibility of simulating different types of virtual quantum systems with different properties with the IBM Q simulator, and that they can be simulated fairly accurately, irrespective of the contributions of environmental noise and CNOT noise. We show a decoupling procedure that lets decouples the NV centres from their environment and allows them to maintain their entanglement in the process, which could be a way toward their usage in scalable and fault tolerant information processing. Finally, we recommend further studies of virtual systems on the platform. 

\smallbreak
\section*{Supplementary information}
 
In the supplemental material we look at the mechanisms involved in entanglement and simulations involving them, additional more general spin qubit simulations, fidelity simulations which relate to the first virtual system proposed and simulations of a dynamical decoupling protocol that could be incorporated as part of the virtual system to improve the resulting coherence times and fidelities. 

\smallbreak
\section*{Acknowledgments}
We thank S. Mukhin (from MISIS), Y. Hardy and P. Madonsela for discussion. SB thanks CSIR-NLC, the URC Wits and National Research Foundation (SA) for the BRICS multilateral program for funding. SB also acknowledges financial support from the Ministry of Education and Science of the Russian Federation in the framework of the Increased Competitiveness Program of NUST MISiS (grant No. K $3-2018-043$).

\bibliography{main.bib}

\providecommand{\noopsort}[1]{}\providecommand{\singleletter}[1]{#1}%
\begin{thebibliography}{49}%
\makeatletter
\providecommand \@ifxundefined [1]{%
 \@ifx{#1\undefined}
}%
\providecommand \@ifnum [1]{%
 \ifnum #1\expandafter \@firstoftwo
 \else \expandafter \@secondoftwo
 \fi
}%
\providecommand \@ifx [1]{%
 \ifx #1\expandafter \@firstoftwo
 \else \expandafter \@secondoftwo
 \fi
}%
\providecommand \natexlab [1]{#1}%
\providecommand \enquote  [1]{``#1''}%
\providecommand \bibnamefont  [1]{#1}%
\providecommand \bibfnamefont [1]{#1}%
\providecommand \citenamefont [1]{#1}%
\providecommand \href@noop [0]{\@secondoftwo}%
\providecommand \href [0]{\begingroup \@sanitize@url \@href}%
\providecommand \@href[1]{\@@startlink{#1}\@@href}%
\providecommand \@@href[1]{\endgroup#1\@@endlink}%
\providecommand \@sanitize@url [0]{\catcode `\\12\catcode `\$12\catcode
  `\&12\catcode `\#12\catcode `\^12\catcode `\_12\catcode `\%12\relax}%
\providecommand \@@startlink[1]{}%
\providecommand \@@endlink[0]{}%
\providecommand \url  [0]{\begingroup\@sanitize@url \@url }%
\providecommand \@url [1]{\endgroup\@href {#1}{\urlprefix }}%
\providecommand \urlprefix  [0]{URL }%
\providecommand \Eprint [0]{\href }%
\providecommand \doibase [0]{http://dx.doi.org/}%
\providecommand \selectlanguage [0]{\@gobble}%
\providecommand \bibinfo  [0]{\@secondoftwo}%
\providecommand \bibfield  [0]{\@secondoftwo}%
\providecommand \translation [1]{[#1]}%
\providecommand \BibitemOpen [0]{}%
\providecommand \bibitemStop [0]{}%
\providecommand \bibitemNoStop [0]{.\EOS\space}%
\providecommand \EOS [0]{\spacefactor3000\relax}%
\providecommand \BibitemShut  [1]{\csname bibitem#1\endcsname}%
\let\auto@bib@innerbib\@empty
\bibitem [{\citenamefont {Biamonte}\ \emph {et~al.}(2017)\citenamefont
  {Biamonte}, \citenamefont {Wittek}, \citenamefont {Pancotti}, \citenamefont
  {Rebentrost}, \citenamefont {Wiebe},\ and\ \citenamefont
  {Lloyd}}]{Biamonte2017}%
  \BibitemOpen
  \bibfield  {author} {\bibinfo {author} {\bibfnamefont {J.}~\bibnamefont
  {Biamonte}}, \bibinfo {author} {\bibfnamefont {P.}~\bibnamefont {Wittek}},
  \bibinfo {author} {\bibfnamefont {N.}~\bibnamefont {Pancotti}}, \bibinfo
  {author} {\bibfnamefont {P.}~\bibnamefont {Rebentrost}}, \bibinfo {author}
  {\bibfnamefont {N.}~\bibnamefont {Wiebe}}, \ and\ \bibinfo {author}
  {\bibfnamefont {S.}~\bibnamefont {Lloyd}},\ }\href
  {https://doi.org/10.1038/nature23474 http://10.0.4.14/nature23474} {\bibfield
   {journal} {\bibinfo  {journal} {Nature}\ }\textbf {\bibinfo {volume}
  {549}},\ \bibinfo {pages} {195} (\bibinfo {year} {2017})}\BibitemShut
  {NoStop}%
\bibitem [{\citenamefont {Georgescu}, \citenamefont {Ashhab},\ and\
  \citenamefont {Nori}(2014)}]{RevModPhys.86.153}%
  \BibitemOpen
  \bibfield  {author} {\bibinfo {author} {\bibfnamefont {I.~M.}\ \bibnamefont
  {Georgescu}}, \bibinfo {author} {\bibfnamefont {S.}~\bibnamefont {Ashhab}}, \
  and\ \bibinfo {author} {\bibfnamefont {F.}~\bibnamefont {Nori}},\ }\href
  {\doibase 10.1103/RevModPhys.86.153} {\bibfield  {journal} {\bibinfo
  {journal} {Rev. Mod. Phys.}\ }\textbf {\bibinfo {volume} {86}},\ \bibinfo
  {pages} {153} (\bibinfo {year} {2014})}\BibitemShut {NoStop}%
\bibitem [{\citenamefont {Lesovik}\ \emph {et~al.}(2019)\citenamefont
  {Lesovik}, \citenamefont {Sadovskyy}, \citenamefont {Suslov}, \citenamefont
  {Lebedev},\ and\ \citenamefont {Vinokur}}]{Lesovik2019}%
  \BibitemOpen
  \bibfield  {author} {\bibinfo {author} {\bibfnamefont {G.~B.}\ \bibnamefont
  {Lesovik}}, \bibinfo {author} {\bibfnamefont {I.~A.}\ \bibnamefont
  {Sadovskyy}}, \bibinfo {author} {\bibfnamefont {M.~V.}\ \bibnamefont
  {Suslov}}, \bibinfo {author} {\bibfnamefont {A.~V.}\ \bibnamefont {Lebedev}},
  \ and\ \bibinfo {author} {\bibfnamefont {V.~M.}\ \bibnamefont {Vinokur}},\
  }\href {\doibase 10.1038/s41598-019-40765-6} {\bibfield  {journal} {\bibinfo
  {journal} {Sci. Rep}\ }\textbf {\bibinfo {volume} {9}},\ \bibinfo {pages}
  {4396} (\bibinfo {year} {2019})}\BibitemShut {NoStop}%
\bibitem [{\citenamefont {Zhukov}\ \emph {et~al.}(2017)\citenamefont {Zhukov},
  , \citenamefont {Pogosov},\ and\ \citenamefont
  {Lozovik}}]{Zhukov2017ModelingDO}%
  \BibitemOpen
  \bibfield  {author} {\bibinfo {author} {\bibfnamefont {A.~A.}\ \bibnamefont
  {Zhukov}}, , \bibinfo {author} {\bibfnamefont {W.~V.}\ \bibnamefont
  {Pogosov}}, \ and\ \bibinfo {author} {\bibfnamefont {Y.~E.}\ \bibnamefont
  {Lozovik}}\ }(\bibinfo {year} {2017})\ \Eprint
  {http://arxiv.org/abs/1710.09659v1} {arXiv:1710.09659v1 [quant-ph]}
  \BibitemShut {NoStop}%
\bibitem [{\citenamefont {Garcia-Martin}\ and\ \citenamefont
  {Sierra}(2018)}]{GarciaMartin2018FiveET}%
  \BibitemOpen
  \bibfield  {author} {\bibinfo {author} {\bibfnamefont {D.}~\bibnamefont
  {Garcia-Martin}}\ and\ \bibinfo {author} {\bibfnamefont {G.}~\bibnamefont
  {Sierra}}\ }(\bibinfo {year} {2018})\ \Eprint
  {http://arxiv.org/abs/1712.05642v4} {arXiv:1712.05642v4 [quant-ph]}
  \BibitemShut {NoStop}%
\bibitem [{\citenamefont {Steudtner}\ and\ \citenamefont
  {Wehner}(2019)}]{PhysRevA.99.022308}%
  \BibitemOpen
  \bibfield  {author} {\bibinfo {author} {\bibfnamefont {M.}~\bibnamefont
  {Steudtner}}\ and\ \bibinfo {author} {\bibfnamefont {S.}~\bibnamefont
  {Wehner}},\ }\href {\doibase 10.1103/PhysRevA.99.022308} {\bibfield
  {journal} {\bibinfo  {journal} {Phys. Rev. A}\ }\textbf {\bibinfo {volume}
  {99}},\ \bibinfo {pages} {022308} (\bibinfo {year} {2019})}\BibitemShut
  {NoStop}%
\bibitem [{\citenamefont {Raeisi}, \citenamefont {Wiebe},\ and\ \citenamefont
  {Sanders}(2012)}]{Raeisi2012}%
  \BibitemOpen
  \bibfield  {author} {\bibinfo {author} {\bibfnamefont {S.}~\bibnamefont
  {Raeisi}}, \bibinfo {author} {\bibfnamefont {N.}~\bibnamefont {Wiebe}}, \
  and\ \bibinfo {author} {\bibfnamefont {B.~C.}\ \bibnamefont {Sanders}},\
  }\href {\doibase 10.1088/1367-2630/14/10/103017} {\bibfield  {journal}
  {\bibinfo  {journal} {New J. Phys.}\ }\textbf {\bibinfo {volume} {14}},\
  \bibinfo {pages} {103017} (\bibinfo {year} {2012})}\BibitemShut {NoStop}%
\bibitem [{\citenamefont {Smith}\ \emph {et~al.}(2019)\citenamefont {Smith},
  \citenamefont {Kim}, \citenamefont {Pollmann},\ and\ \citenamefont
  {Knolle}}]{Smith:2019mek}%
  \BibitemOpen
  \bibfield  {author} {\bibinfo {author} {\bibfnamefont {A.}~\bibnamefont
  {Smith}}, \bibinfo {author} {\bibfnamefont {M.~S.}\ \bibnamefont {Kim}},
  \bibinfo {author} {\bibfnamefont {F.}~\bibnamefont {Pollmann}}, \ and\
  \bibinfo {author} {\bibfnamefont {J.}~\bibnamefont {Knolle}},\ }\href@noop {}
  {\  (\bibinfo {year} {2019})},\ \Eprint {http://arxiv.org/abs/1906.06343}
  {arXiv:1906.06343 [quant-ph]} \BibitemShut {NoStop}%
\bibitem [{\citenamefont {Hegade}, \citenamefont {Behera},\ and\ \citenamefont
  {Panigrahi}(2017)}]{Hegade2017ExperimentalDO}%
  \BibitemOpen
  \bibfield  {author} {\bibinfo {author} {\bibfnamefont {N.~N.}\ \bibnamefont
  {Hegade}}, \bibinfo {author} {\bibfnamefont {B.~K.}\ \bibnamefont {Behera}},
  \ and\ \bibinfo {author} {\bibfnamefont {P.~K.}\ \bibnamefont {Panigrahi}}\
  }(\bibinfo {year} {2017})\ \Eprint {http://arxiv.org/abs/1712.07326v4}
  {arXiv:1712.07326v4 [quant-ph]} \BibitemShut {NoStop}%
\bibitem [{\citenamefont {Kapil}, \citenamefont {Behera},\ and\ \citenamefont
  {Panigrahi}(2018)}]{Kapil2018QuantumSO}%
  \BibitemOpen
  \bibfield  {author} {\bibinfo {author} {\bibfnamefont {M.}~\bibnamefont
  {Kapil}}, \bibinfo {author} {\bibfnamefont {B.~K.}\ \bibnamefont {Behera}}, \
  and\ \bibinfo {author} {\bibfnamefont {P.~K.}\ \bibnamefont {Panigrahi}}\
  }(\bibinfo {year} {2018})\ \Eprint {http://arxiv.org/abs/1807.00521v2}
  {arXiv:1807.00521v2 [quant-ph]} \BibitemShut {NoStop}%
\bibitem [{\citenamefont {Gerritsma}\ \emph {et~al.}(2010)\citenamefont
  {Gerritsma}, \citenamefont {Kirchmair}, \citenamefont {Z{\"a}hringer},
  \citenamefont {Solano}, \citenamefont {Blatt},\ and\ \citenamefont
  {Roos}}]{Gerritsma2010QuantumSO}%
  \BibitemOpen
  \bibfield  {author} {\bibinfo {author} {\bibfnamefont {R.}~\bibnamefont
  {Gerritsma}}, \bibinfo {author} {\bibfnamefont {G.}~\bibnamefont
  {Kirchmair}}, \bibinfo {author} {\bibfnamefont {F.}~\bibnamefont
  {Z{\"a}hringer}}, \bibinfo {author} {\bibfnamefont {E.~C.~M.}\ \bibnamefont
  {Solano}}, \bibinfo {author} {\bibfnamefont {R.}~\bibnamefont {Blatt}}, \
  and\ \bibinfo {author} {\bibfnamefont {C.~F.}\ \bibnamefont {Roos}},\ }\href
  {https://doi.org/10.1038/nature08688 http://10.0.4.14/nature08688} {\bibfield
   {journal} {\bibinfo  {journal} {Nature}\ }\textbf {\bibinfo {volume}
  {463}},\ \bibinfo {pages} {68} (\bibinfo {year} {2010})}\BibitemShut
  {NoStop}%
\bibitem [{\citenamefont {Wen}\ \emph {et~al.}(2019)\citenamefont {Wen},
  \citenamefont {Zheng}, \citenamefont {Kong}, \citenamefont {Wei},
  \citenamefont {Xin},\ and\ \citenamefont {Long}}]{PhysRevA.99.062122}%
  \BibitemOpen
  \bibfield  {author} {\bibinfo {author} {\bibfnamefont {J.}~\bibnamefont
  {Wen}}, \bibinfo {author} {\bibfnamefont {C.}~\bibnamefont {Zheng}}, \bibinfo
  {author} {\bibfnamefont {X.}~\bibnamefont {Kong}}, \bibinfo {author}
  {\bibfnamefont {S.}~\bibnamefont {Wei}}, \bibinfo {author} {\bibfnamefont
  {T.}~\bibnamefont {Xin}}, \ and\ \bibinfo {author} {\bibfnamefont
  {G.}~\bibnamefont {Long}},\ }\href {\doibase 10.1103/PhysRevA.99.062122}
  {\bibfield  {journal} {\bibinfo  {journal} {Phys. Rev. A}\ }\textbf {\bibinfo
  {volume} {99}},\ \bibinfo {pages} {062122} (\bibinfo {year}
  {2019})}\BibitemShut {NoStop}%
\bibitem [{\citenamefont {Chang}\ \emph {et~al.}(2014)\citenamefont {Chang},
  \citenamefont {Xing}, \citenamefont {Zhang}, \citenamefont {Liu},
  \citenamefont {Jiang}, \citenamefont {Li}, \citenamefont {Gu}, \citenamefont
  {Long},\ and\ \citenamefont {Pan}}]{Chang}%
  \BibitemOpen
  \bibfield  {author} {\bibinfo {author} {\bibfnamefont {Y.-C.}\ \bibnamefont
  {Chang}}, \bibinfo {author} {\bibfnamefont {J.}~\bibnamefont {Xing}},
  \bibinfo {author} {\bibfnamefont {F.-H.}\ \bibnamefont {Zhang}}, \bibinfo
  {author} {\bibfnamefont {G.-Q.}\ \bibnamefont {Liu}}, \bibinfo {author}
  {\bibfnamefont {Q.-Q.}\ \bibnamefont {Jiang}}, \bibinfo {author}
  {\bibfnamefont {W.-X.}\ \bibnamefont {Li}}, \bibinfo {author} {\bibfnamefont
  {C.-Z.}\ \bibnamefont {Gu}}, \bibinfo {author} {\bibfnamefont {G.-L.}\
  \bibnamefont {Long}}, \ and\ \bibinfo {author} {\bibfnamefont {X.-Y.}\
  \bibnamefont {Pan}},\ }\href {\doibase 10.1063/1.4885772} {\bibfield
  {journal} {\bibinfo  {journal} {Appl. Phys. Lett.}\ }\textbf {\bibinfo
  {volume} {104}},\ \bibinfo {pages} {262403} (\bibinfo {year}
  {2014})}\BibitemShut {NoStop}%
\bibitem [{\citenamefont {Yan}\ \emph {et~al.}(2019)\citenamefont {Yan},
  \citenamefont {Zhang}, \citenamefont {Gong} \emph {et~al.}}]{Yan753}%
  \BibitemOpen
  \bibfield  {author} {\bibinfo {author} {\bibfnamefont {Z.}~\bibnamefont
  {Yan}}, \bibinfo {author} {\bibfnamefont {Y.-R.}\ \bibnamefont {Zhang}},
  \bibinfo {author} {\bibfnamefont {M.}~\bibnamefont {Gong}},  \emph {et~al.},\
  }\href {\doibase 10.1126/science.aaw1611} {\bibfield  {journal} {\bibinfo
  {journal} {Science}\ }\textbf {\bibinfo {volume} {364}},\ \bibinfo {pages}
  {753} (\bibinfo {year} {2019})}\BibitemShut {NoStop}%
\bibitem [{\citenamefont {Moll}\ \emph {et~al.}(2018)\citenamefont {Moll},
  \citenamefont {Barkoutsos}, \citenamefont {Bishop} \emph
  {et~al.}}]{Moll_2018}%
  \BibitemOpen
  \bibfield  {author} {\bibinfo {author} {\bibfnamefont {N.}~\bibnamefont
  {Moll}}, \bibinfo {author} {\bibfnamefont {P.}~\bibnamefont {Barkoutsos}},
  \bibinfo {author} {\bibfnamefont {L.~S.}\ \bibnamefont {Bishop}},  \emph
  {et~al.},\ }\href {\doibase 10.1088/2058-9565/aab822} {\bibfield  {journal}
  {\bibinfo  {journal} {Quantum Sci Technol}\ }\textbf {\bibinfo {volume}
  {3}},\ \bibinfo {pages} {030503} (\bibinfo {year} {2018})}\BibitemShut
  {NoStop}%
\bibitem [{\citenamefont {Chao}\ and\ \citenamefont
  {Reichardt}(2018)}]{Chao2018}%
  \BibitemOpen
  \bibfield  {author} {\bibinfo {author} {\bibfnamefont {R.}~\bibnamefont
  {Chao}}\ and\ \bibinfo {author} {\bibfnamefont {B.~W.}\ \bibnamefont
  {Reichardt}},\ }\href {\doibase 10.1038/s41534-018-0085-z} {\bibfield
  {journal} {\bibinfo  {journal} {npj Quantum Inf.}\ }\textbf {\bibinfo
  {volume} {4}},\ \bibinfo {pages} {42} (\bibinfo {year} {2018})}\BibitemShut
  {NoStop}%
\bibitem [{\citenamefont {Kurizki}\ \emph {et~al.}(2015)\citenamefont
  {Kurizki}, \citenamefont {Bertet}, \citenamefont {Kubo}, \citenamefont
  {M{\o}lmer}, \citenamefont {Petrosyan}, \citenamefont {Rabl},\ and\
  \citenamefont {Schmiedmayer}}]{Kurizki3866}%
  \BibitemOpen
  \bibfield  {author} {\bibinfo {author} {\bibfnamefont {G.}~\bibnamefont
  {Kurizki}}, \bibinfo {author} {\bibfnamefont {P.}~\bibnamefont {Bertet}},
  \bibinfo {author} {\bibfnamefont {Y.}~\bibnamefont {Kubo}}, \bibinfo {author}
  {\bibfnamefont {K.}~\bibnamefont {M{\o}lmer}}, \bibinfo {author}
  {\bibfnamefont {D.}~\bibnamefont {Petrosyan}}, \bibinfo {author}
  {\bibfnamefont {P.}~\bibnamefont {Rabl}}, \ and\ \bibinfo {author}
  {\bibfnamefont {J.}~\bibnamefont {Schmiedmayer}},\ }\href {\doibase
  10.1073/pnas.1419326112} {\bibfield  {journal} {\bibinfo  {journal} {PNAS}\
  }\textbf {\bibinfo {volume} {112}},\ \bibinfo {pages} {3866} (\bibinfo {year}
  {2015})}\BibitemShut {NoStop}%
\bibitem [{\citenamefont {Kubo}\ \emph {et~al.}(2011)\citenamefont {Kubo},
  \citenamefont {Grezes}, \citenamefont {Dewes} \emph
  {et~al.}}]{PhysRevLett.107.220501}%
  \BibitemOpen
  \bibfield  {author} {\bibinfo {author} {\bibfnamefont {Y.}~\bibnamefont
  {Kubo}}, \bibinfo {author} {\bibfnamefont {C.}~\bibnamefont {Grezes}},
  \bibinfo {author} {\bibfnamefont {A.}~\bibnamefont {Dewes}},  \emph
  {et~al.},\ }\href {\doibase 10.1103/PhysRevLett.107.220501} {\bibfield
  {journal} {\bibinfo  {journal} {Phys. Rev. Lett.}\ }\textbf {\bibinfo
  {volume} {107}},\ \bibinfo {pages} {220501} (\bibinfo {year}
  {2011})}\BibitemShut {NoStop}%
\bibitem [{\citenamefont {Marcos}\ \emph {et~al.}(2010)\citenamefont {Marcos},
  \citenamefont {Wubs}, \citenamefont {Taylor}, \citenamefont {Aguado},
  \citenamefont {Lukin},\ and\ \citenamefont
  {S\o{}rensen}}]{PhysRevLett.105.210501}%
  \BibitemOpen
  \bibfield  {author} {\bibinfo {author} {\bibfnamefont {D.}~\bibnamefont
  {Marcos}}, \bibinfo {author} {\bibfnamefont {M.}~\bibnamefont {Wubs}},
  \bibinfo {author} {\bibfnamefont {J.~M.}\ \bibnamefont {Taylor}}, \bibinfo
  {author} {\bibfnamefont {R.}~\bibnamefont {Aguado}}, \bibinfo {author}
  {\bibfnamefont {M.~D.}\ \bibnamefont {Lukin}}, \ and\ \bibinfo {author}
  {\bibfnamefont {A.~S.}\ \bibnamefont {S\o{}rensen}},\ }\href {\doibase
  10.1103/PhysRevLett.105.210501} {\bibfield  {journal} {\bibinfo  {journal}
  {Phys. Rev. Lett.}\ }\textbf {\bibinfo {volume} {105}},\ \bibinfo {pages}
  {210501} (\bibinfo {year} {2010})}\BibitemShut {NoStop}%
\bibitem [{\citenamefont {Qiu}\ \emph {et~al.}(2014)\citenamefont {Qiu},
  \citenamefont {Xiong}, \citenamefont {Tian},\ and\ \citenamefont
  {You}}]{Qiu2014AHQ}%
  \BibitemOpen
  \bibfield  {author} {\bibinfo {author} {\bibfnamefont {Y.}~\bibnamefont
  {Qiu}}, \bibinfo {author} {\bibfnamefont {W.}~\bibnamefont {Xiong}}, \bibinfo
  {author} {\bibfnamefont {L.}~\bibnamefont {Tian}}, \ and\ \bibinfo {author}
  {\bibfnamefont {J.~Q.}\ \bibnamefont {You}}\ }(\bibinfo {year} {2014})\
  \Eprint {http://arxiv.org/abs/1401.3095v2} {arXiv:1401.3095v2 [quant-ph]}
  \BibitemShut {NoStop}%
\bibitem [{\citenamefont {van~der Sar}\ \emph {et~al.}(2012)\citenamefont
  {van~der Sar}, \citenamefont {Wang}, \citenamefont {Blok}, \citenamefont
  {Bernien}, \citenamefont {Taminiau}, \citenamefont {Toyli}, \citenamefont
  {Lidar}, \citenamefont {Awschalom}, \citenamefont {Hanson},\ and\
  \citenamefont {Dobrovitski}}]{VanderSar2012}%
  \BibitemOpen
  \bibfield  {author} {\bibinfo {author} {\bibfnamefont {T.}~\bibnamefont
  {van~der Sar}}, \bibinfo {author} {\bibfnamefont {Z.~H.}\ \bibnamefont
  {Wang}}, \bibinfo {author} {\bibfnamefont {M.~S.}\ \bibnamefont {Blok}},
  \bibinfo {author} {\bibfnamefont {H.}~\bibnamefont {Bernien}}, \bibinfo
  {author} {\bibfnamefont {T.~H.}\ \bibnamefont {Taminiau}}, \bibinfo {author}
  {\bibfnamefont {D.~M.}\ \bibnamefont {Toyli}}, \bibinfo {author}
  {\bibfnamefont {D.~A.}\ \bibnamefont {Lidar}}, \bibinfo {author}
  {\bibfnamefont {D.~D.}\ \bibnamefont {Awschalom}}, \bibinfo {author}
  {\bibfnamefont {R.}~\bibnamefont {Hanson}}, \ and\ \bibinfo {author}
  {\bibfnamefont {V.~V.}\ \bibnamefont {Dobrovitski}},\ }\href
  {https://doi.org/10.1038/nature10900 http://10.0.4.14/nature10900
  https://www.nature.com/articles/nature10900{\#}supplementary-information}
  {\bibfield  {journal} {\bibinfo  {journal} {Nature}\ }\textbf {\bibinfo
  {volume} {484}},\ \bibinfo {pages} {82} (\bibinfo {year} {2012})}\BibitemShut
  {NoStop}%
\bibitem [{\citenamefont {Taminiau}\ \emph {et~al.}(2014)\citenamefont
  {Taminiau}, \citenamefont {Cramer}, \citenamefont {van~der Sar},
  \citenamefont {Dobrovitski},\ and\ \citenamefont {Hanson}}]{Taminiau2014}%
  \BibitemOpen
  \bibfield  {author} {\bibinfo {author} {\bibfnamefont {T.~H.}\ \bibnamefont
  {Taminiau}}, \bibinfo {author} {\bibfnamefont {J.}~\bibnamefont {Cramer}},
  \bibinfo {author} {\bibfnamefont {T.}~\bibnamefont {van~der Sar}}, \bibinfo
  {author} {\bibfnamefont {V.~V.}\ \bibnamefont {Dobrovitski}}, \ and\ \bibinfo
  {author} {\bibfnamefont {R.}~\bibnamefont {Hanson}},\ }\href
  {https://doi.org/10.1038/nnano.2014.2 http://10.0.4.14/nnano.2014.2
  https://www.nature.com/articles/nnano.2014.2{\#}supplementary-information}
  {\bibfield  {journal} {\bibinfo  {journal} {Nat. Nanotechnol}\ }\textbf
  {\bibinfo {volume} {9}},\ \bibinfo {pages} {171} (\bibinfo {year}
  {2014})}\BibitemShut {NoStop}%
\bibitem [{\citenamefont {Zu}\ \emph {et~al.}(2014)\citenamefont {Zu},
  \citenamefont {Wang}, \citenamefont {He}, \citenamefont {Zhang},
  \citenamefont {Dai}, \citenamefont {Wang},\ and\ \citenamefont
  {Duan}}]{Zu2014}%
  \BibitemOpen
  \bibfield  {author} {\bibinfo {author} {\bibfnamefont {C.}~\bibnamefont
  {Zu}}, \bibinfo {author} {\bibfnamefont {W.-B.}\ \bibnamefont {Wang}},
  \bibinfo {author} {\bibfnamefont {L.}~\bibnamefont {He}}, \bibinfo {author}
  {\bibfnamefont {W.-G.}\ \bibnamefont {Zhang}}, \bibinfo {author}
  {\bibfnamefont {C.-Y.}\ \bibnamefont {Dai}}, \bibinfo {author} {\bibfnamefont
  {F.}~\bibnamefont {Wang}}, \ and\ \bibinfo {author} {\bibfnamefont {L.-M.}\
  \bibnamefont {Duan}},\ }\href {https://doi.org/10.1038/nature13729
  http://10.0.4.14/nature13729} {\bibfield  {journal} {\bibinfo  {journal}
  {Nature}\ }\textbf {\bibinfo {volume} {514}},\ \bibinfo {pages} {72}
  (\bibinfo {year} {2014})}\BibitemShut {NoStop}%
\bibitem [{\citenamefont {Nagata}\ \emph {et~al.}(2018)\citenamefont {Nagata},
  \citenamefont {Kuramitani}, \citenamefont {Sekiguchi},\ and\ \citenamefont
  {Kosaka}}]{Nagata2018}%
  \BibitemOpen
  \bibfield  {author} {\bibinfo {author} {\bibfnamefont {K.}~\bibnamefont
  {Nagata}}, \bibinfo {author} {\bibfnamefont {K.}~\bibnamefont {Kuramitani}},
  \bibinfo {author} {\bibfnamefont {Y.}~\bibnamefont {Sekiguchi}}, \ and\
  \bibinfo {author} {\bibfnamefont {H.}~\bibnamefont {Kosaka}},\ }\href
  {\doibase 10.1038/s41467-018-05664-w} {\bibfield  {journal} {\bibinfo
  {journal} {Nat. Commun}\ }\textbf {\bibinfo {volume} {9}},\ \bibinfo {pages}
  {3227} (\bibinfo {year} {2018})}\BibitemShut {NoStop}%
\bibitem [{\citenamefont {Feng}, \citenamefont {Xu},\ and\ \citenamefont
  {Long}(2013)}]{PhysRevLett.110.190501}%
  \BibitemOpen
  \bibfield  {author} {\bibinfo {author} {\bibfnamefont {G.}~\bibnamefont
  {Feng}}, \bibinfo {author} {\bibfnamefont {G.}~\bibnamefont {Xu}}, \ and\
  \bibinfo {author} {\bibfnamefont {G.}~\bibnamefont {Long}},\ }\href
  {https://link.aps.org/doi/10.1103/PhysRevLett.110.190501} {\bibfield
  {journal} {\bibinfo  {journal} {Phys. Rev. Lett.}\ }\textbf {\bibinfo
  {volume} {110}},\ \bibinfo {pages} {190501} (\bibinfo {year}
  {2013})}\BibitemShut {NoStop}%
\bibitem [{\citenamefont {Li}, \citenamefont {Liu},\ and\ \citenamefont
  {Long}(2017)}]{Li2017}%
  \BibitemOpen
  \bibfield  {author} {\bibinfo {author} {\bibfnamefont {H.}~\bibnamefont
  {Li}}, \bibinfo {author} {\bibfnamefont {Y.}~\bibnamefont {Liu}}, \ and\
  \bibinfo {author} {\bibfnamefont {G.}~\bibnamefont {Long}},\ }\href {\doibase
  10.1007/s11433-017-9058-7} {\bibfield  {journal} {\bibinfo  {journal} {SCI
  CHINA PHYS MECH.}\ }\textbf {\bibinfo {volume} {60}},\ \bibinfo {pages}
  {080311} (\bibinfo {year} {2017})}\BibitemShut {NoStop}%
\bibitem [{\citenamefont {Zhou}\ \emph {et~al.}(2017)\citenamefont {Zhou},
  \citenamefont {Liu}, \citenamefont {Hong},\ and\ \citenamefont
  {Xue}}]{Zhou2017}%
  \BibitemOpen
  \bibfield  {author} {\bibinfo {author} {\bibfnamefont {J.}~\bibnamefont
  {Zhou}}, \bibinfo {author} {\bibfnamefont {B.}~\bibnamefont {Liu}}, \bibinfo
  {author} {\bibfnamefont {Z.}~\bibnamefont {Hong}}, \ and\ \bibinfo {author}
  {\bibfnamefont {Z.}~\bibnamefont {Xue}},\ }\href {\doibase
  10.1007/s11433-017-9119-8} {\bibfield  {journal} {\bibinfo  {journal} {SCI
  CHINA PHYS MECH.}\ }\textbf {\bibinfo {volume} {61}},\ \bibinfo {pages}
  {010312} (\bibinfo {year} {2017})}\BibitemShut {NoStop}%
\bibitem [{\citenamefont {Devitt}(2016)}]{PhysRevA.94.032329}%
  \BibitemOpen
  \bibfield  {author} {\bibinfo {author} {\bibfnamefont {S.~J.}\ \bibnamefont
  {Devitt}},\ }\href {\doibase 10.1103/PhysRevA.94.032329} {\bibfield
  {journal} {\bibinfo  {journal} {Phys. Rev. A}\ }\textbf {\bibinfo {volume}
  {94}},\ \bibinfo {pages} {032329} (\bibinfo {year} {2016})}\BibitemShut
  {NoStop}%
\bibitem [{\citenamefont {Ju}\ \emph {et~al.}(2014)\citenamefont {Ju},
  \citenamefont {Lei}, \citenamefont {Xu}, \citenamefont {Culcer},
  \citenamefont {Zhang},\ and\ \citenamefont {Du}}]{PhysRevB.89.045432}%
  \BibitemOpen
  \bibfield  {author} {\bibinfo {author} {\bibfnamefont {C.}~\bibnamefont
  {Ju}}, \bibinfo {author} {\bibfnamefont {C.}~\bibnamefont {Lei}}, \bibinfo
  {author} {\bibfnamefont {X.}~\bibnamefont {Xu}}, \bibinfo {author}
  {\bibfnamefont {D.}~\bibnamefont {Culcer}}, \bibinfo {author} {\bibfnamefont
  {Z.}~\bibnamefont {Zhang}}, \ and\ \bibinfo {author} {\bibfnamefont
  {J.}~\bibnamefont {Du}},\ }\href
  {https://link.aps.org/doi/10.1103/PhysRevB.89.045432} {\bibfield  {journal}
  {\bibinfo  {journal} {Phys. Rev. B}\ }\textbf {\bibinfo {volume} {89}},\
  \bibinfo {pages} {045432} (\bibinfo {year} {2014})}\BibitemShut {NoStop}%
\bibitem [{\citenamefont {Hanson}, \citenamefont {Gywat},\ and\ \citenamefont
  {Awschalom}(2006)}]{PhysRevB.74.161203}%
  \BibitemOpen
  \bibfield  {author} {\bibinfo {author} {\bibfnamefont {R.}~\bibnamefont
  {Hanson}}, \bibinfo {author} {\bibfnamefont {O.}~\bibnamefont {Gywat}}, \
  and\ \bibinfo {author} {\bibfnamefont {D.~D.}\ \bibnamefont {Awschalom}},\
  }\href {\doibase 10.1103/PhysRevB.74.161203} {\bibfield  {journal} {\bibinfo
  {journal} {Phys. Rev. B}\ }\textbf {\bibinfo {volume} {74}},\ \bibinfo
  {pages} {161203} (\bibinfo {year} {2006})}\BibitemShut {NoStop}%
\bibitem [{\citenamefont {Jing}\ and\ \citenamefont {Wu}(2018)}]{Jing2018}%
  \BibitemOpen
  \bibfield  {author} {\bibinfo {author} {\bibfnamefont {J.}~\bibnamefont
  {Jing}}\ and\ \bibinfo {author} {\bibfnamefont {L.-A.}\ \bibnamefont {Wu}},\
  }\href {\doibase 10.1038/s41598-018-19977-9} {\bibfield  {journal} {\bibinfo
  {journal} {Sci. Rep}\ }\textbf {\bibinfo {volume} {8}},\ \bibinfo {pages}
  {1471} (\bibinfo {year} {2018})}\BibitemShut {NoStop}%
\bibitem [{\citenamefont {Childress}\ \emph {et~al.}(2006)\citenamefont
  {Childress}, \citenamefont {Gurudev~Dutt}, \citenamefont {Taylor},
  \citenamefont {Zibrov}, \citenamefont {Jelezko}, \citenamefont {Wrachtrup},
  \citenamefont {Hemmer},\ and\ \citenamefont {Lukin}}]{Childress281}%
  \BibitemOpen
  \bibfield  {author} {\bibinfo {author} {\bibfnamefont {L.}~\bibnamefont
  {Childress}}, \bibinfo {author} {\bibfnamefont {M.~V.}\ \bibnamefont
  {Gurudev~Dutt}}, \bibinfo {author} {\bibfnamefont {J.~M.}\ \bibnamefont
  {Taylor}}, \bibinfo {author} {\bibfnamefont {A.~S.}\ \bibnamefont {Zibrov}},
  \bibinfo {author} {\bibfnamefont {F.}~\bibnamefont {Jelezko}}, \bibinfo
  {author} {\bibfnamefont {J.}~\bibnamefont {Wrachtrup}}, \bibinfo {author}
  {\bibfnamefont {P.~R.}\ \bibnamefont {Hemmer}}, \ and\ \bibinfo {author}
  {\bibfnamefont {M.~D.}\ \bibnamefont {Lukin}},\ }\href {\doibase
  10.1126/science.1131871} {\bibfield  {journal} {\bibinfo  {journal}
  {Science}\ }\textbf {\bibinfo {volume} {314}},\ \bibinfo {pages} {281}
  (\bibinfo {year} {2006})}\BibitemShut {NoStop}%
\bibitem [{\citenamefont {Jacques}\ \emph {et~al.}(2009)\citenamefont
  {Jacques}, \citenamefont {Neumann}, \citenamefont {Beck}, \citenamefont
  {Markham}, \citenamefont {Twitchen}, \citenamefont {Meijer}, \citenamefont
  {Kaiser}, \citenamefont {Balasubramanian}, \citenamefont {Jelezko},\ and\
  \citenamefont {Wrachtrup}}]{PhysRevLett.102.057403}%
  \BibitemOpen
  \bibfield  {author} {\bibinfo {author} {\bibfnamefont {V.}~\bibnamefont
  {Jacques}}, \bibinfo {author} {\bibfnamefont {P.}~\bibnamefont {Neumann}},
  \bibinfo {author} {\bibfnamefont {J.}~\bibnamefont {Beck}}, \bibinfo {author}
  {\bibfnamefont {M.}~\bibnamefont {Markham}}, \bibinfo {author} {\bibfnamefont
  {D.}~\bibnamefont {Twitchen}}, \bibinfo {author} {\bibfnamefont
  {J.}~\bibnamefont {Meijer}}, \bibinfo {author} {\bibfnamefont
  {F.}~\bibnamefont {Kaiser}}, \bibinfo {author} {\bibfnamefont
  {G.}~\bibnamefont {Balasubramanian}}, \bibinfo {author} {\bibfnamefont
  {F.}~\bibnamefont {Jelezko}}, \ and\ \bibinfo {author} {\bibfnamefont
  {J.}~\bibnamefont {Wrachtrup}},\ }\href {\doibase
  10.1103/PhysRevLett.102.057403} {\bibfield  {journal} {\bibinfo  {journal}
  {Phys. Rev. Lett.}\ }\textbf {\bibinfo {volume} {102}},\ \bibinfo {pages}
  {057403} (\bibinfo {year} {2009})}\BibitemShut {NoStop}%
\bibitem [{\citenamefont {Lamata}(2017)}]{Lamata2017}%
  \BibitemOpen
  \bibfield  {author} {\bibinfo {author} {\bibfnamefont {L.}~\bibnamefont
  {Lamata}},\ }\href {https://doi.org/10.1038/srep43768
  http://10.0.4.14/srep43768} {\bibfield  {journal} {\bibinfo  {journal} {Sci.
  Rep}\ }\textbf {\bibinfo {volume} {7}},\ \bibinfo {pages} {43768} (\bibinfo
  {year} {2017})}\BibitemShut {NoStop}%
\bibitem [{\citenamefont {Bar-Gill}\ \emph {et~al.}(2013)\citenamefont
  {Bar-Gill}, \citenamefont {Pham}, \citenamefont {Jarmola}, \citenamefont
  {Budker},\ and\ \citenamefont {Walsworth}}]{Bar-Gill2013}%
  \BibitemOpen
  \bibfield  {author} {\bibinfo {author} {\bibfnamefont {N.}~\bibnamefont
  {Bar-Gill}}, \bibinfo {author} {\bibfnamefont {L.~M.}\ \bibnamefont {Pham}},
  \bibinfo {author} {\bibfnamefont {A.}~\bibnamefont {Jarmola}}, \bibinfo
  {author} {\bibfnamefont {D.}~\bibnamefont {Budker}}, \ and\ \bibinfo {author}
  {\bibfnamefont {R.~L.}\ \bibnamefont {Walsworth}},\ }\href
  {https://doi.org/10.1038/nature10900 http://10.0.4.14/nature10900
  https://www.nature.com/articles/nature10900{\#}supplementary-information}
  {\bibfield  {journal} {\bibinfo  {journal} {Nat. Commun}\ }\textbf {\bibinfo
  {volume} {4}},\ \bibinfo {pages} {1743} (\bibinfo {year} {2013})}\BibitemShut
  {NoStop}%
\bibitem [{\citenamefont {Bauch}\ \emph {et~al.}(2018)\citenamefont {Bauch},
  \citenamefont {Hart}, \citenamefont {Schloss}, \citenamefont {Turner},
  \citenamefont {Barry}, \citenamefont {Kehayias}, \citenamefont {Singh},\ and\
  \citenamefont {Walsworth}}]{PhysRevX.8.031025}%
  \BibitemOpen
  \bibfield  {author} {\bibinfo {author} {\bibfnamefont {E.}~\bibnamefont
  {Bauch}}, \bibinfo {author} {\bibfnamefont {C.~A.}\ \bibnamefont {Hart}},
  \bibinfo {author} {\bibfnamefont {J.~M.}\ \bibnamefont {Schloss}}, \bibinfo
  {author} {\bibfnamefont {M.~J.}\ \bibnamefont {Turner}}, \bibinfo {author}
  {\bibfnamefont {J.~F.}\ \bibnamefont {Barry}}, \bibinfo {author}
  {\bibfnamefont {P.}~\bibnamefont {Kehayias}}, \bibinfo {author}
  {\bibfnamefont {S.}~\bibnamefont {Singh}}, \ and\ \bibinfo {author}
  {\bibfnamefont {R.~L.}\ \bibnamefont {Walsworth}},\ }\href
  {https://link.aps.org/doi/10.1103/PhysRevX.8.031025} {\bibfield  {journal}
  {\bibinfo  {journal} {Phys. Rev. X}\ }\textbf {\bibinfo {volume} {8}},\
  \bibinfo {pages} {031025} (\bibinfo {year} {2018})}\BibitemShut {NoStop}%
\bibitem [{\citenamefont {Herbschleb}\ \emph {et~al.}(2019)\citenamefont
  {Herbschleb}, \citenamefont {Kato}, \citenamefont {Maruyama}, \citenamefont
  {Danjo}, \citenamefont {Makino}, \citenamefont {Yamasaki}, \citenamefont
  {Ohki}, \citenamefont {Hayashi}, \citenamefont {Morishita}, \citenamefont
  {Fujiwara},\ and\ \citenamefont {Mizuochi}}]{Herbschleb2019}%
  \BibitemOpen
  \bibfield  {author} {\bibinfo {author} {\bibfnamefont {E.~D.}\ \bibnamefont
  {Herbschleb}}, \bibinfo {author} {\bibfnamefont {H.}~\bibnamefont {Kato}},
  \bibinfo {author} {\bibfnamefont {Y.}~\bibnamefont {Maruyama}}, \bibinfo
  {author} {\bibfnamefont {T.}~\bibnamefont {Danjo}}, \bibinfo {author}
  {\bibfnamefont {T.}~\bibnamefont {Makino}}, \bibinfo {author} {\bibfnamefont
  {S.}~\bibnamefont {Yamasaki}}, \bibinfo {author} {\bibfnamefont
  {I.}~\bibnamefont {Ohki}}, \bibinfo {author} {\bibfnamefont {K.}~\bibnamefont
  {Hayashi}}, \bibinfo {author} {\bibfnamefont {H.}~\bibnamefont {Morishita}},
  \bibinfo {author} {\bibfnamefont {M.}~\bibnamefont {Fujiwara}}, \ and\
  \bibinfo {author} {\bibfnamefont {N.}~\bibnamefont {Mizuochi}},\ }\href
  {\doibase 10.1038/s41467-019-11776-8} {\bibfield  {journal} {\bibinfo
  {journal} {Nat. Commun.}\ }\textbf {\bibinfo {volume} {10}},\ \bibinfo
  {pages} {3766} (\bibinfo {year} {2019})}\BibitemShut {NoStop}%
\bibitem [{\citenamefont {Viola}, \citenamefont {Knill},\ and\ \citenamefont
  {Lloyd}(1999)}]{PhysRevLett.82.2417}%
  \BibitemOpen
  \bibfield  {author} {\bibinfo {author} {\bibfnamefont {L.}~\bibnamefont
  {Viola}}, \bibinfo {author} {\bibfnamefont {E.}~\bibnamefont {Knill}}, \ and\
  \bibinfo {author} {\bibfnamefont {S.}~\bibnamefont {Lloyd}},\ }\href
  {\doibase 10.1103/PhysRevLett.82.2417} {\bibfield  {journal} {\bibinfo
  {journal} {Phys. Rev. Lett.}\ }\textbf {\bibinfo {volume} {82}},\ \bibinfo
  {pages} {2417} (\bibinfo {year} {1999})}\BibitemShut {NoStop}%
\bibitem [{\citenamefont {Pokharel}\ \emph {et~al.}(2018)\citenamefont
  {Pokharel}, \citenamefont {Anand}, \citenamefont {Fortman},\ and\
  \citenamefont {Lidar}}]{PhysRevLett.121.220502}%
  \BibitemOpen
  \bibfield  {author} {\bibinfo {author} {\bibfnamefont {B.}~\bibnamefont
  {Pokharel}}, \bibinfo {author} {\bibfnamefont {N.}~\bibnamefont {Anand}},
  \bibinfo {author} {\bibfnamefont {B.}~\bibnamefont {Fortman}}, \ and\
  \bibinfo {author} {\bibfnamefont {D.~A.}\ \bibnamefont {Lidar}},\ }\href
  {\doibase 10.1103/PhysRevLett.121.220502} {\bibfield  {journal} {\bibinfo
  {journal} {Phys. Rev. Lett.}\ }\textbf {\bibinfo {volume} {121}},\ \bibinfo
  {pages} {220502} (\bibinfo {year} {2018})}\BibitemShut {NoStop}%
\bibitem [{\citenamefont {Wang}\ \emph {et~al.}(2012)\citenamefont {Wang},
  \citenamefont {de~Lange}, \citenamefont {Rist\`e}, \citenamefont {Hanson},\
  and\ \citenamefont {Dobrovitski}}]{PhysRevB.85.155204}%
  \BibitemOpen
  \bibfield  {author} {\bibinfo {author} {\bibfnamefont {Z.-H.}\ \bibnamefont
  {Wang}}, \bibinfo {author} {\bibfnamefont {G.}~\bibnamefont {de~Lange}},
  \bibinfo {author} {\bibfnamefont {D.}~\bibnamefont {Rist\`e}}, \bibinfo
  {author} {\bibfnamefont {R.}~\bibnamefont {Hanson}}, \ and\ \bibinfo {author}
  {\bibfnamefont {V.~V.}\ \bibnamefont {Dobrovitski}},\ }\href {\doibase
  10.1103/PhysRevB.85.155204} {\bibfield  {journal} {\bibinfo  {journal} {Phys.
  Rev. B}\ }\textbf {\bibinfo {volume} {85}},\ \bibinfo {pages} {155204}
  (\bibinfo {year} {2012})}\BibitemShut {NoStop}%
\bibitem [{\citenamefont {de~Lange}\ \emph {et~al.}(2010)\citenamefont
  {de~Lange}, \citenamefont {Wang}, \citenamefont {Rist{\`e}}, \citenamefont
  {Dobrovitski},\ and\ \citenamefont {Hanson}}]{deLange60}%
  \BibitemOpen
  \bibfield  {author} {\bibinfo {author} {\bibfnamefont {G.}~\bibnamefont
  {de~Lange}}, \bibinfo {author} {\bibfnamefont {Z.~H.}\ \bibnamefont {Wang}},
  \bibinfo {author} {\bibfnamefont {D.}~\bibnamefont {Rist{\`e}}}, \bibinfo
  {author} {\bibfnamefont {V.~V.}\ \bibnamefont {Dobrovitski}}, \ and\ \bibinfo
  {author} {\bibfnamefont {R.}~\bibnamefont {Hanson}},\ }\href {\doibase
  10.1126/science.1192739} {\bibfield  {journal} {\bibinfo  {journal}
  {Science}\ }\textbf {\bibinfo {volume} {330}},\ \bibinfo {pages} {60}
  (\bibinfo {year} {2010})}\BibitemShut {NoStop}%
\bibitem [{\citenamefont {Hanson}\ \emph {et~al.}(2008)\citenamefont {Hanson},
  \citenamefont {Dobrovitski}, \citenamefont {Feiguin}, \citenamefont {Gywat},\
  and\ \citenamefont {Awschalom}}]{Hanson352}%
  \BibitemOpen
  \bibfield  {author} {\bibinfo {author} {\bibfnamefont {R.}~\bibnamefont
  {Hanson}}, \bibinfo {author} {\bibfnamefont {V.~V.}\ \bibnamefont
  {Dobrovitski}}, \bibinfo {author} {\bibfnamefont {A.~E.}\ \bibnamefont
  {Feiguin}}, \bibinfo {author} {\bibfnamefont {O.}~\bibnamefont {Gywat}}, \
  and\ \bibinfo {author} {\bibfnamefont {D.~D.}\ \bibnamefont {Awschalom}},\
  }\href {\doibase 10.1126/science.1155400} {\bibfield  {journal} {\bibinfo
  {journal} {Science}\ }\textbf {\bibinfo {volume} {320}},\ \bibinfo {pages}
  {352} (\bibinfo {year} {2008})}\BibitemShut {NoStop}%
\bibitem [{\citenamefont {Yan}\ \emph {et~al.}(2016)\citenamefont {Yan},
  \citenamefont {Gustavsson}, \citenamefont {Kamal} \emph {et~al.}}]{Yan2016}%
  \BibitemOpen
  \bibfield  {author} {\bibinfo {author} {\bibfnamefont {F.}~\bibnamefont
  {Yan}}, \bibinfo {author} {\bibfnamefont {S.}~\bibnamefont {Gustavsson}},
  \bibinfo {author} {\bibfnamefont {A.}~\bibnamefont {Kamal}},  \emph
  {et~al.},\ }\href {https://doi.org/10.1038/ncomms12964
  http://10.0.4.14/ncomms12964
  https://www.nature.com/articles/ncomms12964{\#}supplementary-information}
  {\bibfield  {journal} {\bibinfo  {journal} {Nat. Commun}\ }\textbf {\bibinfo
  {volume} {7}},\ \bibinfo {pages} {12964} (\bibinfo {year}
  {2016})}\BibitemShut {NoStop}%
\bibitem [{\citenamefont {Mr{\'o}zek}\ \emph {et~al.}(2015)\citenamefont
  {Mr{\'o}zek}, \citenamefont {Rudnicki}, \citenamefont {Kehayias},
  \citenamefont {Jarmola}, \citenamefont {Budker},\ and\ \citenamefont
  {Gawlik}}]{Mrozek2015}%
  \BibitemOpen
  \bibfield  {author} {\bibinfo {author} {\bibfnamefont {M.}~\bibnamefont
  {Mr{\'o}zek}}, \bibinfo {author} {\bibfnamefont {D.}~\bibnamefont
  {Rudnicki}}, \bibinfo {author} {\bibfnamefont {P.}~\bibnamefont {Kehayias}},
  \bibinfo {author} {\bibfnamefont {A.}~\bibnamefont {Jarmola}}, \bibinfo
  {author} {\bibfnamefont {D.}~\bibnamefont {Budker}}, \ and\ \bibinfo {author}
  {\bibfnamefont {W.}~\bibnamefont {Gawlik}},\ }\href {\doibase
  10.1140/epjqt/s40507-015-0035-z} {\bibfield  {journal} {\bibinfo  {journal}
  {EPJ Quantum Technol.}\ }\textbf {\bibinfo {volume} {2}},\ \bibinfo {pages}
  {22} (\bibinfo {year} {2015})}\BibitemShut {NoStop}%
\bibitem [{\citenamefont {Schwartz}\ \emph {et~al.}(2018)\citenamefont
  {Schwartz}, \citenamefont {Scheuer}, \citenamefont {Tratzmiller} \emph
  {et~al.}}]{Schwartzeaat8978}%
  \BibitemOpen
  \bibfield  {author} {\bibinfo {author} {\bibfnamefont {I.}~\bibnamefont
  {Schwartz}}, \bibinfo {author} {\bibfnamefont {J.}~\bibnamefont {Scheuer}},
  \bibinfo {author} {\bibfnamefont {B.}~\bibnamefont {Tratzmiller}},  \emph
  {et~al.},\ }\href {\doibase 10.1126/sciadv.aat8978} {\bibfield  {journal}
  {\bibinfo  {journal} {Sci. Adv}\ }\textbf {\bibinfo {volume} {4}} (\bibinfo
  {year} {2018}),\ 10.1126/sciadv.aat8978}\BibitemShut {NoStop}%
\bibitem [{\citenamefont {Song}\ \emph {et~al.}(2015)\citenamefont {Song},
  \citenamefont {Yin}, \citenamefont {Yang}, \citenamefont {Zhu}, \citenamefont
  {Zhou},\ and\ \citenamefont {Feng}}]{Song2015}%
  \BibitemOpen
  \bibfield  {author} {\bibinfo {author} {\bibfnamefont {W.-l.}\ \bibnamefont
  {Song}}, \bibinfo {author} {\bibfnamefont {Z.-q.}\ \bibnamefont {Yin}},
  \bibinfo {author} {\bibfnamefont {W.-l.}\ \bibnamefont {Yang}}, \bibinfo
  {author} {\bibfnamefont {X.-b.}\ \bibnamefont {Zhu}}, \bibinfo {author}
  {\bibfnamefont {F.}~\bibnamefont {Zhou}}, \ and\ \bibinfo {author}
  {\bibfnamefont {M.}~\bibnamefont {Feng}},\ }\href
  {https://doi.org/10.1038/srep07755 http://10.0.4.14/srep07755} {\bibfield
  {journal} {\bibinfo  {journal} {Sci. Rep}\ }\textbf {\bibinfo {volume} {5}},\
  \bibinfo {pages} {7755} (\bibinfo {year} {2015})}\BibitemShut {NoStop}%
\bibitem [{\citenamefont {Maleki}\ and\ \citenamefont
  {Zheltikov}(2018)}]{Maleki:18}%
  \BibitemOpen
  \bibfield  {author} {\bibinfo {author} {\bibfnamefont {Y.}~\bibnamefont
  {Maleki}}\ and\ \bibinfo {author} {\bibfnamefont {A.~M.}\ \bibnamefont
  {Zheltikov}},\ }\href {\doibase 10.1364/OE.26.017849} {\bibfield  {journal}
  {\bibinfo  {journal} {Opt. Express}\ }\textbf {\bibinfo {volume} {26}},\
  \bibinfo {pages} {17849} (\bibinfo {year} {2018})}\BibitemShut {NoStop}%
\bibitem [{\citenamefont {Lei}\ \emph {et~al.}(2017)\citenamefont {Lei},
  \citenamefont {Peng}, \citenamefont {Ju}, \citenamefont {Yung},\ and\
  \citenamefont {Du}}]{Lei2017}%
  \BibitemOpen
  \bibfield  {author} {\bibinfo {author} {\bibfnamefont {C.}~\bibnamefont
  {Lei}}, \bibinfo {author} {\bibfnamefont {S.}~\bibnamefont {Peng}}, \bibinfo
  {author} {\bibfnamefont {C.}~\bibnamefont {Ju}}, \bibinfo {author}
  {\bibfnamefont {M.-H.}\ \bibnamefont {Yung}}, \ and\ \bibinfo {author}
  {\bibfnamefont {J.}~\bibnamefont {Du}},\ }\href {\doibase
  10.1038/s41598-017-12280-z} {\bibfield  {journal} {\bibinfo  {journal} {Sci.
  Rep}\ }\textbf {\bibinfo {volume} {7}},\ \bibinfo {pages} {11937} (\bibinfo
  {year} {2017})}\BibitemShut {NoStop}%
\bibitem [{\citenamefont {Liu}, \citenamefont {You},\ and\ \citenamefont
  {Hou}(2016)}]{Liu2016a}%
  \BibitemOpen
  \bibfield  {author} {\bibinfo {author} {\bibfnamefont {Y.}~\bibnamefont
  {Liu}}, \bibinfo {author} {\bibfnamefont {J.}~\bibnamefont {You}}, \ and\
  \bibinfo {author} {\bibfnamefont {Q.}~\bibnamefont {Hou}},\ }\href
  {https://doi.org/10.1038/srep21775 http://10.0.4.14/srep21775} {\bibfield
  {journal} {\bibinfo  {journal} {Sci. Rep}\ }\textbf {\bibinfo {volume} {6}},\
  \bibinfo {pages} {21775} (\bibinfo {year} {2016})}\BibitemShut {NoStop}%
\end{thebibliography}%

\end{document}